\documentclass[aps,prl,reprint,superscriptaddress,showpacs]{revtex4-1}

\usepackage{graphicx,amsmath,amssymb,ulem}
\usepackage{verbatim}
\usepackage{ragged2e}
\usepackage{color}
\begin{document}
\title{Broadband bright twin beams and their up-conversion}

\author{Maria V.~Chekhova}
\affiliation{Max Planck Institute for the Science of Light, Staudtstra\ss{}e 2, 91058 Erlangen, Germany}
\affiliation{University of Erlangen-N\"urnberg, Staudtstra\ss{}e 7/B2, 91058 Erlangen, Germany}
\affiliation{Department of Physics, M.~V.~Lomonosov Moscow State University, Leninskie Gory, 119991 Moscow, Russia}

\author{Semen Germanskiy}
\affiliation{Department of Physics, M.~V.~Lomonosov Moscow State University, Leninskie Gory, 119991 Moscow, Russia}
\affiliation{Helmholz Zentrum Dresden Rossendorf, Bautzner Landstr. 400, 01328 Dresden, Germany}

\author{Dmitri B.~Horoshko}
\affiliation{Univ. Lille, CNRS, UMR 8523—PhLAM F-59000 Lille, France }
\affiliation{B. I. Stepanov Institute of Physics, NASB, Nezavisimosti Ave. 68, 220072 Minsk, Belarus}

\author{Galiya Kh.~Kitaeva}
\affiliation{Department of Physics, M.~V.~Lomonosov Moscow State University, Leninskie Gory, 119991 Moscow, Russia}

\author{Mikhail I.~Kolobov}
\affiliation{Univ. Lille, CNRS, UMR 8523—PhLAM F-59000 Lille, France }

\author{Gerd Leuchs}
\affiliation{Max Planck Institute for the Science of Light, Staudtstra\ss{}e 2, 91058 Erlangen, Germany}
\affiliation{University of Erlangen-N\"urnberg, Staudtstra\ss{}e 7/B2, 91058 Erlangen, Germany}

\author{Chris R.~Phillips}
\affiliation{Department of Physics, ETH Zurich, CH-8093 Zurich, Switzerland}

\author{Pavel A.~Prudkovskii}
\affiliation{Department of Physics, M.~V.~Lomonosov Moscow State University, Leninskie Gory, 119991 Moscow, Russia}

\begin{abstract}
We report on the observation of broadband ($40$ THz) bright twin beams through high-gain parametric down-conversion in an aperiodically poled lithium niobate crystal. The output photon number is shown to scale exponentially with the pump power and not with the pump amplitude, as in homogeneous crystals. Photon-number correlations and the number of frequency/temporal modes are assessed by spectral covariance measurements.  By using sum-frequency generation on the surface of a non-phasematched crystal, we measure a cross-correlation peak with the temporal width $90$ fs.
\end{abstract}

\maketitle

Within the broad subject of quantum state engineering, there is a lot of interest to the generation of ultrabroadband nonclassical light, in particular ultrabroadband or even single-cycle photon pairs (biphotons)~\cite{Shaked2014,Harris,Sensarn,Nasr}. A broad spectrum in combination with a narrow correlation bandwidth means a high degree of frequency/time entanglement~\cite{Brida2009, Horoshko2012} and can be used for high-dimensional quantum information encoding~\cite{Bessire}. Moreover, the larger the bandwidth of an entangled source, the faster quantum communication it can provide~\cite{Dayan2005}. Another motivation for engineering such sources is due to nonlinear optical effects with nonclassical light~\cite{Klyshko1982,GeaBanacloche,Georgiades} and clock synchronization~\cite{Valencia}.

Biphotons with ultrabroad spectrum and a high degree of frequency entanglement can be obtained through parametric down-conversion (PDC) under frequency-degenerate phase matching at wavelengths with low group velocity dispersion~\cite{Strekalov2005,ODonnell2007,Shaked2014}, in a long crystal pumped by short pulses~\cite{Brida2009}, by taking advantage of the whole frequency-angle spectrum of PDC~\cite{Dayan2004,Katamadze2015}, or simply by using a very thin nonlinear crystal provided that the pumping is strong enough~\cite{Katamadze2013}.

Another method to achieve an ultrabroad spectrum is to use aperiodically poled crystals~\cite{Harris,Nasr,Sensarn,Fejer2008a,Fejer2008b,Horoshko2013,Horoshko2017,Phillips2014}. It enables not only broadening, but also engineering of the spectral shape in an arbitrary way, limited of course by the technology level. Another considerable advantage of this method is that it is principally lossless, in contrast to ones that involve diffractive or phase elements.

At the same time, the spectral broadening achieved through aperiodic poling is inhomogeneous, in the sense that photons at different wavelengths are generated at different positions in the crystal. For this reason, a phase chirp appears in the two-photon spectral amplitude, and reaching short correlation times requires its removal. Under inhomogeneous spectral broadening, the frequency dependence of the phase contains the squared phase mismatch both in the low-gain~\cite{Harris} and the high-gain regimes~\cite{Horoshko2017}. Thus the phase is a fourth-order polynomial of the frequency even in the simplest case of a quadratic dispersion dependence. For not very broad spectra the dominant quadratic phase chirp can be eliminated by using, for instance, a material with large group-velocity dispersion (GVD)~\cite{Harris,Brida}. However, this strategy does not work if the spectrum is broadened to values exceeding 10 THz~\cite{Horoshko2017}. In this case, elimination of the phase chirp requires either active compensation or a special design of the crystal~\cite{Horoshko2017}.

At strong pumping, PDC produces not biphotons but bright squeezed vacuum~\cite{Jedrkiewicz2004,Bondani,Brida1,Agafonov}: radiation with strong photon-number correlations and quadrature squeezing. Due to its high degree of photon-number entanglement~\cite{Chekhova2015}, this quantum state of light has a high quantum information capacity, which can be further increased due to a large number of frequency modes.
The ultimate bandwidth limit of the whole optical octave is attained for ``single-cycle squeezed light'' considered in ~\cite{Horoshko2013}.

\begin{figure}[htbp]
\centering
\includegraphics[width=0.95\linewidth]{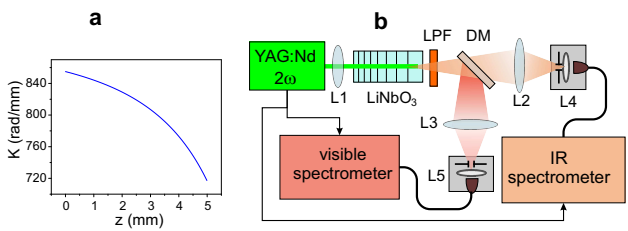}
\caption{The design of the aperiodic poling (a) and the setup for the study of spectral correlations (b). }
\label{fig:setup_spectrum}
\end{figure}

In our experiment we generate bright squeezed vacuum in an aperiodically poled  $5$ mm long sample of lithium niobate (LiNbO$_3$) crystal doped with MgO. The samples were produced by Gooch and Housego~\cite{Gooch}. According to the design, the inverse grating vector should vary along the crystal as a squared hyperbolic function~\cite{Horoshko2017},
\begin{equation}
K(z)=-\frac{\alpha}{4(2-z/L)^2}+\beta,
\label{eq:K(z)}
\end{equation}
with $\alpha=735$ rad/mm, $\beta=901$ rad/mm, and $z$ is the longitudinal coordinate (Fig.~\ref{fig:setup_spectrum}a). Such a design provides compensation for the the third-order and fourth-order terms in the phase chirp without any additional optical elements. With the grating vector as shown in Fig.~\ref{fig:setup_spectrum}a, type-0 phase matching condition is satisfied for pumping at 532 nm, the signal and idler wavelengths being around 790 nm and 1600 nm, respectively.

The setup for the study of the spectral properties is shown in Fig.~\ref{fig:setup_spectrum}b. The pump is the second-harmonic radiation of a Nd:YAG laser, producing 18-ps full-width at half maximum (FWHM) pulses at a wavelength of 532 nm, focused into the LiNbO$_3$ crystal by lens L1 with the focal distance $500$ mm. The repetition rate is 1 kHz. The pump power in the crystal is up to $25$ mW, which corresponds to the peak intensity $0.51$ GW/cm$^2$, based on a focused beam size of $170\,\mu$m $1/e^2$ radius measured in air without the crystal.

 After the crystal, the pump is cut off by a long-pass filter (LPF) and the signal and idler beams are separated by a dichroic mirror DM. In each beam, there is a lens (L2, L3), in whose focal plane an aperture selects nearly collinear emission: $5$ mrad and $2.5$ mrad for the idler and signal beams, respectively. After each aperture, the radiation is collected by lenses L4, L5 into a multimode fiber and fed into a spectrometer:  AvaSpec-ULS3648 for the signal radiation and AvaSpec-NIRS12 for the idler radiation. The spectrometers are gated synchronously with the laser pulse.

\begin{figure}[htbp]
\centering
\includegraphics[width=0.95\linewidth]{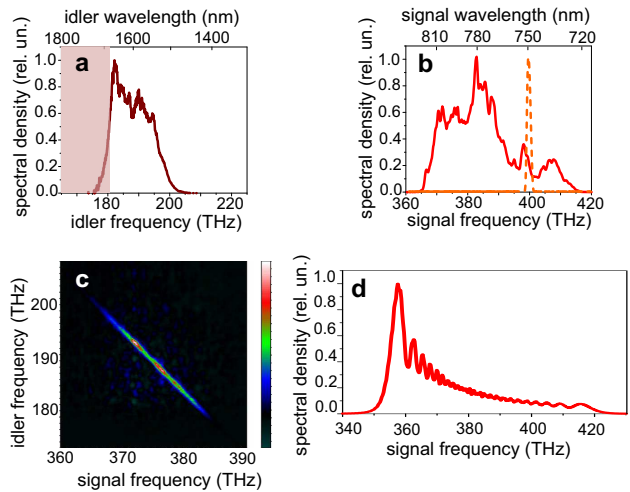}
\caption{Spectra of the signal and idler radiation emitted in the collinear direction (a,b) and the covariance describing the spectral correlations between them (c). Dashed orange line in (b) shows the spectrum of a periodically poled sample. Panel (d) shows the spectrum calculated according to the theory of Ref~\cite{Horoshko2017} with $\nu_0=0.1$.}
\label{fig:spectral}
\end{figure}

The resulting spectra obtained for the signal and idler beams under pumping with the average power $10$ mW are shown in Fig.~\ref{fig:spectral}a,b, respectively. The spectral sensitivities of both signal and idler arms of the setup were calibrated using the method described in Ref.~\cite{Lemieux2016}, and the resulting spectra were corrected for this sensitivity. However, the NIR arm has zero sensitivity for wavelengths above $1650$ nm (shown by shading); this is why the idler spectrum is cut on the `red' side. At the same time, the spectrum of the signal beam is fully registered by the visible spectrometer, and the spectral width is $40$ THz, or $90$ nm. The peaks observed in the spectra are caused by the non-idealities of the domain structure.

The spectrum expected for the ideal grating as shown in Fig.~\ref{fig:setup_spectrum}a can be calculated using the approach of Ref.~\cite{Horoshko2017}. We introduce the positive-frequency field for PDC light as
\begin{equation}
E^{(+)}(t,z)=e^{-i\omega_0t}\int_{-\omega_0}^{\omega_0}\hat{b}(\Omega,z)e^{i\{k(\omega_0+\Omega)z-\Omega t\} }d\Omega,
\label{eq:field}
\end{equation}
where $\omega_0=\omega_p/2$ is half the pump frequency and $\hat{b}(\Omega,z)$ the PDC photon annihilation operator at frequency $\omega_0+\Omega$.  Its evolution is given by the Bogolyubov transformation,
\begin{equation}
\hat{b}(\Omega,z)=A(\Omega,z)\hat{b}(\Omega,0)+B(\Omega,z)\hat{b}^\dagger(-\Omega,0).
\label{eq:Bogolyubov}
\end{equation}
The evolution of $A(\Omega,z)$ and $B(\Omega,z)$ is described by the system of equations
\begin{eqnarray}
\frac{\partial A(\Omega,z)}{\partial z}&=&igB^*(\Omega,z)e^{i\Delta(\Omega)z-i\int_0^zK(z')dz'},\nonumber\\
\frac{\partial B^*(\Omega,z)}{\partial z}&=&-ig^*A(\Omega,z)e^{-i\Delta(\Omega)z+i\int_0^zK(z')dz'},
\label{eq:system}
\end{eqnarray}
where $\Delta(\Omega)=k_p-k(\omega_0+\Omega)-k(\omega_0-\Omega)$ is the wavevector mismatch and $g$ the parametric interaction coefficient involving the pump field and the quadratic susceptibility. By numerically integrating this system, we obtain the spectrum of the signal beam $S(\Omega)=|B(\Omega,L)|^2$.  Fig.~\ref{fig:spectral}d shows this spectrum for the amplification coefficient $\nu_0=|g|^2L/|K(0)-K(L)|=0.1$. Compared to the theoretical prediction, the experimental spectrum is cut on the `blue' side and especially on the `red' side, probably due to the deviations of the $K(z)$ dependence from Eq.(\ref{eq:K(z)}).

The observed $40$ THz spectral width is far from being a record (compare, for instance, with Refs.~\cite{Shaked2014,Katamadze2015}), but is quite large for non-degenerate PDC.  For comparison, panel (b) shows the signal spectrum obtained for a LiNbO$_3$ crystal of the same length, poled with the constant grating vector $774$ rad/mm (dashed orange line). The width is $1.3$ THz, or $2.5$ nm, which shows that the aperiodic poling leads to the spectral broadening by a factor of 36.

To characterize the frequency correlations between the signal and idler beams, we take a set of $3000$ single-pulse spectra and measure the covariance of signal and idler photon numbers as a function of frequencies $\omega_s,\,\omega_i$~\cite{Spasibko2012,Finger2016},
\begin{equation}
\mathrm{Cov}(\omega_s,\omega_i)=\langle N(\omega_s)N(\omega_i)\rangle-\langle N(\omega_s)\rangle\langle N(\omega_i)\rangle.
\label{eq:Cov}
\end{equation}
This distribution is shown in panel c of Fig.~\ref{fig:spectral} and looks as a narrow stripe, which is an evidence of strong frequency correlations. Although the bottom-right part of the stripe is not visible due to the `blindness' of the IR spectrometer in this range, one can still infer the number of modes using the ratio between the total spectral width $\Delta\omega$ and the width $\delta\omega$ of the covariance distribution cross-section~\cite{Brida2009},
\begin{equation}
R=\Delta\omega/\delta\omega.
\label{eq:R}
\end{equation}
The ratio (\ref{eq:R}) is $40$, which is only the lower boundary of the number of modes, because Eq.~(\ref{eq:R}) does not take into account possible additional entanglement in the phase of the two-photon amplitude.

\begin{figure}[htbp]
\centering
\includegraphics[width=0.7\linewidth]{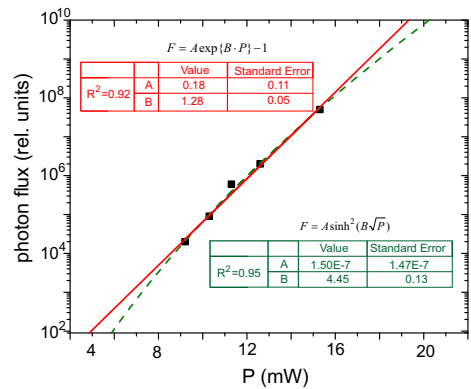}
\caption{Photon flux measured within a band of $5$ nm at the wavelength $1600$ nm as a function of the pump power. Red solid and green dashed lines are fits with different dependences.}
\label{fig:gain}
\end{figure}

The parametric gain is found from the dependence of the output signal after narrowband filtering on the pump power~\cite{Agafonov}. In contrast to the case of a homogeneous crystal, in an aperiodically poled one the effective length of nonlinear interaction scales linearly with the pump field amplitude~\cite{Fejer2008a,Horoshko2017}. As a result, the number of photons per mode at high gain depends on the pump power  $P$ as $N=A(e^{BP}-1)$, where $B$ is a parameter depending on the quadratic nonlinearity and its spatial modulation and $A\sim1$. In other words, the parametric gain exponent $G=BP$ is a linear rather than square root function of the pump power. Using this dependence to fit the photon flux within a $5$ nm bandwidth around $1600$ nm as a function of the pump power (Fig.~\ref{fig:gain}), we find that the parametric gain exponent under pumping at $15$ mW is as high as $G=18$, resulting in $5\cdot10^{7}$ photons per mode. The same figure shows a fit by the $A\sinh^2(B\sqrt{P})$ function, typical for high-gain PDC in a homogeneous crystal. The fit results in an unphysically small $A$ coefficient, which demonstrates its invalidity.

The linear rather than square-root dependence of the parametric gain exponent on the pump power has been found in the classical theory of parametric amplification in aperiodically poled crystals \cite{Fejer2008a} and in the quantum theory of squeezed light generation in such crystals \cite{Horoshko2013,Horoshko2017}. It is generally known as the Rosenbluth gain, first obtained for a problem governed by the same differential equation in plasma physics~\cite{Rosenbluth1972}. It can be understood from a simple argument: the down-converted radiation around some frequency is generated in a layer of the crystal phase-matched for this frequency and has a gain proportional to the pump amplitude times the layer width. Theoretical analysis shows \cite{Horoshko2017} that the width of the layer grows linearly with the pump amplitude; thus, the overall dependence of the gain is quadratic in the pump amplitude, or linear in its power.

\begin{figure}[htbp]
\centering
\includegraphics[width=0.95\linewidth]{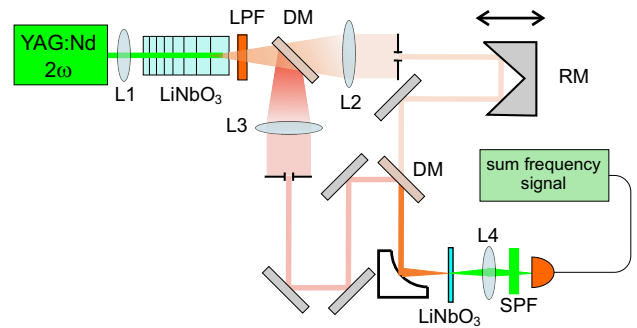}
\caption{The setup for measuring the temporal second-order intensity correlation function. Signal and idler beams, after a delay line, are combined on a dichroic mirror and focused on the surface of a lithium niobate crystal for sum-frequency generation.}
\label{fig:setup_temp}
\end{figure}

Finally, the photon-number correlation time is measured with an ultrafast correlator shown in Fig.~\ref{fig:setup_temp}. The path length difference between the signal and idler beams is scanned by displacing a retroreflecting mirror (RM). After combining the signal and idler beams again on a dichroic mirror (DM), we focus them on the sum-frequency generation (SFG) crystal by means of a gold parabolic mirror.
In contrast to previous work (see, for instance, Ref.~\cite{Dayan2005}) where phase matched SFG was used as a correlator, in our setup the SFG occurs without phase matching, on the surface of a $3$ mm slab of LiNbO$_3$ crystal. Due to the absence of phase matching, the SFG covers a broad spectral range and enables the measurement of ultra-short correlation times. Both signal and idler beams are polarized along the z-axis of the crystal, which provides a high efficiency of nonlinear conversion due to the large value of the quadratic susceptibility $\chi^{(2)}_{zzz}$.  After the SFG, the signal and idler radiation is cut off by a short-pass filter (SPF) while the sum-frequency radiation around 532 nm is collected by lens L4 into a pulsed photon-number integrating detector based on a p-i-n diode. The power of the sum-frequency radiation is measured as a function of the time delay between the signal and the idler beams.

\begin{figure}[htbp]
\centering
\includegraphics[width=0.95\linewidth]{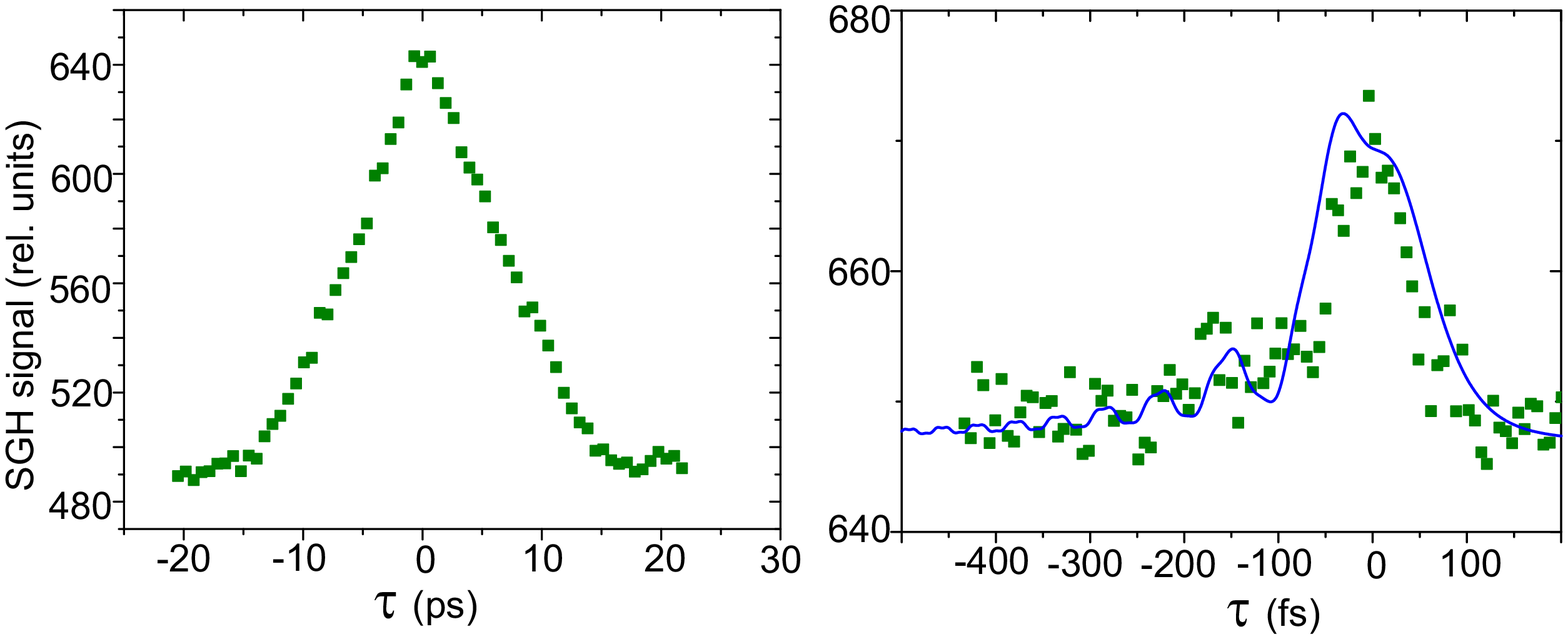}
\caption{Signal of sum frequency generation as a function of the time delay between the signal and idler beams. The left and right panels show the results of scanning with $1$ ps and $10$ fs resolution, respectively. The blue line shows the dependence calculated numerically for gain exponent $G=18$.}
\label{fig:temporal}
\end{figure}

Fig~\ref{fig:temporal} shows the results. Coarse scanning with a step of $1$ ps (left panel) reveals the ‘incoherent’ pedestal with the width determined by the pulse duration. The background is caused by the residual pump and signal radiation. Fine scanning around the top of the pedestal with a step of $10$ fs (right panel) shows a $90$ fs additional peak indicating ultrafast correlations of the twin beams. The reason for the pedestal being pronounced is the high brightness of the twin beams. A similar SHG time dependence was observed in Ref.~\cite{Abram} although with a much broader peak. At the same time, the $90$ fs peak we measure is considerably narrower than the one observed for two-photon light generated by an aperiodically poled crystal~\cite{Sensarn}.

The theoretical dependence of the SFG signal on the delay $\tau$ of the idler field with respect to the signal one can be obtained by writing the total PDC field at the input of the correlator as
\begin{eqnarray}
E^{(+)}(t,L)=e^{-i\omega_0t}\int_0^{\omega_0}[\hat{b}(\Omega,L)e^{i\{k(\omega_0+\Omega)L-\Omega t\}}+\nonumber\\
+\hat{b}(-\Omega,L)e^{i\{k(\omega_0-\Omega)L+\Omega(t+\tau)-\omega_0\tau\}}]d\Omega,
\label{eq:field(t)}
\end{eqnarray}
and by finding the mean intensity of the sum-frequency field,
\begin{equation}\label{eq:SFG}
I_{SFG}\propto\langle\mathrm{vac}|[E^{(-)}(t,L)]^2[E^{(+)}(t,L)]^2|\mathrm{vac}\rangle.
\end{equation}
For a continuous-wave pump, we get the result similar to the one of Refs.~\cite{Horoshko2013,Dayan2007},
\begin{eqnarray}\label{eq:Res_SFG}
I_{SFG}(\tau)\propto8\left(\int_0^{\omega_0}|B(\Omega,L)|^2d\Omega\right)^2+\nonumber\\
+4|\int_0^{\omega_0}A(\Omega,L)B(\Omega,L)e^{i(\Omega\tau-\Delta(\Omega)L)}d\Omega|^2.
\end{eqnarray}
The first term describes the non-coherent background, which in the case of pulsed pump is not constant but repeats the shape of the pulse auto-convolution, while the second term is the coherent component of SHG. Its width depends on the phase $\arg\{A(\Omega,L)B(\Omega,L)\}-\Delta(\Omega)L$. Calculation according to Eq.~(\ref{eq:Res_SFG}) (blue curve in Fig.~\ref{fig:temporal}) shows an asymmetric peak with the width similar to the one observed in the experiment.

The correlation time can be further reduced by compensating the quadratic phase chirp by using additional dispersive material.

In conclusion, we have observed the broadening of the spectrum of bright squeezed vacuum twin beams through the aperiodic poling of the LiNbO$_3$ crystal. Compared to the periodically poled sample of the same length, the spectrum gets broadened by a factor of $36$. This is a relatively large value, taking into account the fact that, theoretically, the spectrum should get narrower with a higher parametric gain.

The output photon number is shown to be exponentially increasing with the pump power according to the Rosenbluth law, typical for aperiodically poled crystals. This is different from the case of homogeneous crystals where the output photon number scales exponentially with the pump amplitude.

Measurement of the spectral covariance of the signal and idler beams revealed a large number of modes, their lower bound equal to $40$.

The use of aperiodic poling with the compensation of higher-order dispersion chirp, even without additional group-velocity dispersion elements, allowed us to observe a $90$ fs broad peak in the temporal correlation function, which we have measured using an ultrafast correlator, based on the non-phasematched sum-frequency generation.

Apart from applications in quantum information, such bright twin beams with ultrashort correlation times will be very useful for nonlinear optics and spectroscopy.

\section*{Funding Information}
This work was supported by the Russian Science Foundation grant No. 17-12-01134.

\end{document}